# What makes a good role model

Jossy Sayir, *Member, IEEE*

*This paper is dedicated to Jim Massey on the occasion of his 75$^{th}$ birthday*

**Abstract**

The role model strategy is introduced as a method for designing an estimator by approaching the output of a superior estimator that has better input observations. This strategy is shown to yield the optimal Bayesian estimator when a Markov condition is fulfilled. Two examples involving simple channels are given to illustrate its use. The strategy is combined with time averaging to construct a statistical model by numerically solving a convex program. The role model strategy was developed in the context of low complexity decoder design for iterative decoding. Potential applications outside the field of communications are discussed.

**Index Terms**

Bayesian estimation, convex programming, low complexity decoding

## I. INTRODUCTION

Picking a role model is a natural way of coping with challenges by striving to emulate a person we admire: children look up to superheroes, musicians emulate Mozart, and information theorists follow Claude E. Shannon. Think of it this way: when faced with a difficult choice, solving your problem on your own may seem like a lonely, daunting prospect. Instead, you seek refuge behind your role model and try to imagine what {Superman/Shannon/your parents/your PhD advisor} would do in your place. But is this a good strategy? Are you better off solving your problem on your own rather than doing what you believe your role model would have done? This paper attempts to answer this question in a probabilistic, Bayesian setting. The main result of the paper is that, if you pick the right role model, the best solution you could obtain on your own is identical to the solution obtained via the "role model" strategy, which might be less tedious.

Despite the philosophical undertone of the previous paragraph, our interest in role models did not originate in mere psychological curiosity. The "role model" terminology is a metaphor that we use to add an intuitive perspective

J. Sayir wrote this paper while on a parental break, with a part-time position at the Telecommunications Research Center in Vienna (ftw.), an institution supported by the Austrian Government and by the City of Vienna within the competence center program COMET, and a visiting position at the Ecole Nationale Supérieure d'Electronique et de ses Applications (ENSEA), Cergy, France. His research at these institutions received the support of the European Commission within the framework of the FP7 Network of Excellence in Wireless COMmunications NEWCOM++ (contract n. 216715) and of the FP7 STREP DAVINCI Codes (contract n. 216203). Although this paper is a private submission documenting non-funded research, these institutions and projects are gratefully acknowledged for their support and for providing part of the motivation for the results presented here.

Submitted on September 8th, 2008

to an otherwise dry statistical question. The motivation for the questions addressed in this paper is the theoretical framework for the design of post-processing functions in iterative decoding using low complexity components as developed in [1], [2], [3]. We believe that the results have applications in other fields where statistical modeling based on derived observations is required and we will show a number of examples to illustrate this. Furthermore, knowing that the "role model" strategy is optimal given certain constraints is not only of academic interest: in the applications considered, it allows us to replace a complex statistical measurement problem by a convex optimization problem with all the convenient baggage of numerical methods that come with it.

In Section II, we will give a mathematical definition of the role model strategy. Section III contains the main theorem stating when the role model strategy is optimal. Section IV shows two examples for which the strategy can be used. Section V discusses potential applications.

## II. THE ROLE MODEL STRATEGY

Let $X$, $Y$ and $Z$ be discrete random variables. In "role model" terminology, $X$ is the quantity in which you are interested, $Y$ is the knowledge available to your role model, and $Z$ is the knowledge available to yourself. Your aim is to use your knowledge $Z$ to make the best possible summary description of $X$. We assume at this point in our analysis that the joint distribution $P_{XYZ}$ is known.

The obvious direct solution of this problem, given the observation $Z = z$, is to compute the *a posteriori* probability distribution $P_{X|Z=z}$ for $X$, which is a description of $X$ that incorporates everthing of statistical interest given the knowledge that $Z = z$. We can write this direct solution as the random variable $S_{\mathsf{d}}(Z)$, which is a function of the random variable $Z$ and whose value when $Z = z$ is

$$S_{\mathsf{d}}(z) = P_{X|Z=z}. \tag{1}$$

Note that $S_{\mathsf{d}}(Z)$ takes values in the space of probability distributions over $X$. This solution is optimal from a Bayesian perspective, and has the added benefit that it is "information lossless" in the sense that

$$I(X; S_{\mathsf{d}}(Z)) = I(X; Z), \tag{2}$$

as was shown in [4].

Throughout our education, we are taught to look carefully at all the "givens" in a problem. In textbook exercises, failure to use one of the "givens" is often a signal that our solution is wrong or sub-optimal. Considered from this perspective, the Bayesian *a posteriori* probability calculation described above is disappointing: the random variable $Y$ does not appear anywhere and the solution makes use of $P_{XZ}$ only, although $P_{XYZ}$ is known. This solution is not specific to the problem and would have been the same if the problem involved only $X$ and $Z$.

To make the solution depend on $Y$, we consider an alternative strategy. If our role model is indeed superior to us with respect to the problem $X$, then we could try to imagine the solution that our role model would have computed given $Y$. Since we do not know the realization $Y = y$, we cannot compute the exact Bayesian solution $P_{X|Y=y}$ that our role model would have computed. We can, however, attempt to produce a distribution $Q_{X|Z=z}$ that gets as close as possible to $P_{X|Y=y}$ in expectation over all possible realizations $y$ of $Y$. Here we have written $Q$ instead of $P$ to stress that this is not generally a true *a posteriori* probability distribution, but rather an arbitrarily chosen function of $Z$ taking values in the space of probability distributions over $X$.

A good measure of "closeness" when it comes to probability distributions is the Kullback-Leibler divergence. For realizations $Z = z$ and $Y = y$, the divergence

$$D(P_{X|Y=y}||Q_{X|Z=z}) = \sum_x P(x|y) \log_2 \frac{P(x|y)}{Q(x|z)} \tag{3}$$

measures how close $Q_{X|Z=z}$ is to $P_{X|Y=y}$, i.e., how close we are to our role model's solution. Averaging the divergence over the possible realizations of $Y$ gives

$$ED(P_{X|Y}||Q_{X|Z=z}) = \sum_y P(y|z) D(P_{X|Y=y}||Q_{X|Z=z}), \tag{4}$$

Averaging the divergence jointly over $Y$ and $Z$ gives

$$ED(P_{X|Y}||Q_{X|Z}) = \sum_y \sum_z P(yz) D(P_{X|Y=y}||Q_{X|Z=z}). \tag{5}$$

We can now state in mathematical terms what we mean by "emulating a role model":

*Definition 1 (The "role model" strategy):* For every possible value $z$ of $Z$, choose the probability distribution $Q_{X|Z=z}$ over $X$ that minimizes $ED(P_{X|Y}||Q_{X|Z=z})$ to be the description $S_{\text{rm}}(z)$ of $X$ given the observation $Z = z$.

Note that since the role model strategy minimizes the expected divergence for every realization $z$ of $Z$, it also does so on average. Thus $S_{\text{rm}}(Z)$ also minimizes $ED(P_{X|Y}||P_{Q|Z})$. In fact, any $Q_{X|Z}$ that minimizes $ED(P_{X|Y}||Q_{X|Z})$ is identical to $S_{\text{rm}}(Z)$ for all $z$ for which $P(z) > 0$. Values of $z$ for which $P(z) = 0$ will not make a difference, so we can replace the minimization for every $z$ by a minimization of the average in our definition.

The variable $Q_{X|Z}$ appears in the denominator inside the logarithm in the definitions of expected divergence (4) and (5). The logarithm is a concave (convex-∩) function and so is its restriction to the convex set of probability distributions for $X$. Since the expected divergence is a negative sum of logarithms of the variables plus a constant, it is a convex (convex-∪) function. Therefore, implementing the role model strategy requires the solution of a convex program.

## III. THE ROLE MODEL THEOREM

We now address the question of how well we do by using the role model strategy defined in the previous section and how different the solution $S_{\text{rm}}(Z)$ is from the direct Bayesian solution $S_{\text{d}}(Z)$. Intuitively, we suspect that emulating your role model is a good idea if your role model knows more about the problem $X$ than you do, but is a bad idea if you know something about the problem that the role model does not.

"Knowing more about" a random variable can be formalized with Markov chains. If the joint distribution $P_{XYZ}$ is such that

$$P(x|yz) = P(x|y) \tag{6}$$

for all possible realizations $x$, $y$ and $z$, then $X$, $Y$ and $Z$ form a Markov chain, denoted $X - Y - Z$. Information theoretically, the Markov condition (6) is equivalent to

$$I(X;Z|Y) = 0. \tag{7}$$

In the role model scenario, if $X$, $Y$ and $Z$ form a Markov chain $X - Y - Z$, then all the knowledge in $Z$ about $X$ has been learned from $Y$.

The following theorem confirms our intuition:

*Theorem 1 (The "role model" theorem):* If $X$, $Y$ and $Z$ form a Markov chain $X - Y - Z$, then

$$\mathrm{E}D(P_{X|Y}||Q_{X|Z}) = H(X|Z) - H(X|Y) + \mathrm{E}D(P_{X|Z}||Q_{X|Z}). \tag{8}$$

In particular,

$$\mathrm{E}D(P_{X|Y}||Q_{X|Z}) \geq H(X|Z) - H(X|Y) \tag{9}$$

with equality if and only if $Q_{X|Z=z} = P_{X|Z=z}$ for all $z$ for which $P(z) > 0$.

*Proof:* Because $X$, $Y$ and $Z$ form a Markov chain $X - Y - Z$, $Z$, $Y$ and $X$ also form a Markov chain $Z - Y - X$. Thus

$$P(x|yz) = P(x|y) \tag{10}$$

so that

$$P(xyz) = P(yz)P(x|y). \tag{11}$$

Substituting into (5) gives

$$\mathrm{E}D(P_{X|Y}||Q_{X|Z}) = \sum_{xyz} P(xyz) \log_2 \frac{P(x|y)}{Q(x|z)} \tag{12}$$

$$= \sum_{xyz} P(xyz) \log_2 \frac{P(x|y)P(x|z)}{Q(x|z)P(x|z)} \tag{13}$$

$$= \sum_{xyz} P(xyz) \log_2 \frac{1}{P(x|z)} P(x|y) \frac{P(x|z)}{Q(x|z)} \tag{14}$$

$$= H(X|Z) - H(X|Y) + \mathrm{E}D(P_{X|Z}||Q_{X|Z}). \tag{15}$$

The inequality (9) now follows from the well-known inequality for expected divergence, cf. [5, 2nd Corollary of Theorem 2.6.3], namely

$$\mathrm{E}D(P_{X|Z}||Q_{X|Z}) \geq 0, \tag{16}$$

with equality if and only if $Q_{X|Z=z} = P_{X|Z=z}$ for every $z$ for which $P(z) > 0$. □

Theorem 1 shows that the role model strategy is optimal, i.e., $S_{\mathrm{rm}}(Z) = S_{\mathrm{d}}(Z)$, when a Markov relation holds between you, your role model, and the problem you are trying to solve. In other words, if your role model is your teacher in the sense that you learned everything you know about $X$ from your role model, then you have chosen an appropriate role model for the problem.

The following theorem treats the general case when $X$, $Y$ and $Z$ do *not necessarily* form a Markov chain:

*Theorem 2:* For any discrete random variables $X$, $Y$, and $Z$,

$$\mathrm{E}D(P_{X|YZ}||Q_{X|Z}) \geq H(X|Z) - H(X|YZ) \tag{17}$$

with equality if and only if, for every $z$ for which $P(z) > 0$, $Q_{X|Z=z}$ and $P_{X|Z=z}$ are identical.

*Proof:* Because

$$I(X;Z|YZ) = 0, \tag{18}$$

it follows that $X$, $(Y, Z)$ and $Z$ form a Markov chain $X - (Y, Z) - Z$. The theorem is obtained by applying Theorem 1 to this Markov chain. □

Theorem 2 shows that in order to achieve Bayesian optimality in the general case, $Q_{X|Z}$ must be chosen to approach $P_{X|YZ}$ in expected divergence, rather than $P_{X|Y}$. When $X$, $Y$ and $Z$ form a Markov chain, then $P_{X|YZ}$ and $P_{X|Y}$ are identical and the two theorems coincide.

## IV. EXAMPLES

We will begin by showing two examples using simple channels for which the Markov condition applies to illustrate the use of the role model strategy and theorem. We will discuss potential applications in the next section.

### A. A first example

Let $X$, $Y$ and $Z$ be connected through Z-channels in the manner illustrated in Figure 1. The resulting channel from $X$ to $Z$ is a Z-channel with crossover probability $3/4$. Let us assume that the channel is driven with a uniform probability distribution $P_X(0) = P_X(1) = 1/2$. In this case, the optimal Bayesian *a posteriori* distribution is

$$P_{X|Z}(0|0) = \frac{P_X(0)P_{Z|X}(0|0)}{P_Z(0)} = 4/7 \tag{19}$$

and similarly $P_{X|Z}(1|0) = 3/7$, $P_{X|Z}(1|1) = 1$, and $P_{X|Z}(0|1) = 0$.

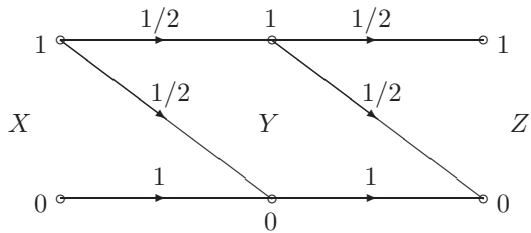

Fig. 1. A cascade of two Z-channels

We now apply the role model approach to this scenario. We define the distribution $Q_{X|Z}$. This distribution has two free parameters that we denote by $q_0 \stackrel{\text{def}}{=} Q_{X|Z}(0|0)$ and $q_1 \stackrel{\text{def}}{=} Q_{X|Z}(1|1)$. We assume that $Z = 0$ was received. We would like to choose $q_0$ to minimize

$$\begin{aligned}
\mathrm{E}D(P_{X|Y}||Q_{X|Z=0}) &= P_{Y|Z}(0|0)\left[P_{X|Y}(0|0)\log_2\frac{P_{X|Y}(0|0)}{q_0} + P_{X|Y}(1|0)\log_2\frac{P_{X|Y}(1|0)}{1-q_0}\right] \\
&\quad + P_{Y|Z}(1|0)\left[P_{X|Y}(0|1)\log_2\frac{P_{X|Y}(0|1)}{q_0} + P_{X|Y}(1|1)\log_2\frac{P_{X|Y}(1|1)}{1-q_0}\right] \\
&= \frac{6}{7}\left[\frac{2}{3}\log_2\frac{2/3}{q_0} + \frac{1}{3}\log_2\frac{1/3}{1-q_0}\right] + \frac{1}{7}\left[0 + 1\log_2\frac{1}{1-q_0}\right] \\
&= -\frac{6}{7}h(1/3) - \frac{4}{7}\log_2 q_0 - \frac{3}{7}\log_2(1-q_0), \qquad (20)
\end{aligned}$$

where $h(.)$ denotes the binary entropy function. Taking the derivative of the expression with respect to $q_0$ and setting it to zero yields $q_0 = 4/7$, which is the same value we obtained using the direct Bayesian approach. Similarly, we get $q_1 = 1$ by applying the same derivation to the case $Z = 1$.

This example shows that the computation resulting from applying the role model approach is different from the computation for the direct Bayesian approach, yet both yield the same solution, in line with the role model theorem. However, so far there appears to be no advantage in replacing the simple computation of (19) by the complicated derivation of (20). The next example will demonstrate an advantage of the role model technique.

## B. A second example

Let us consider the setup in Figure 2, where a binary erasure channel (BEC) with known erasure probability $\delta$ is followed by an unknown channel with ternary input alphabet and binary output alphabet. The channel is driven with uniform inputs, i.e., $P_X(0) = P_X(1) = 1/2$. Note that we now depart from our assumption that $P_{XYZ}$ is known.

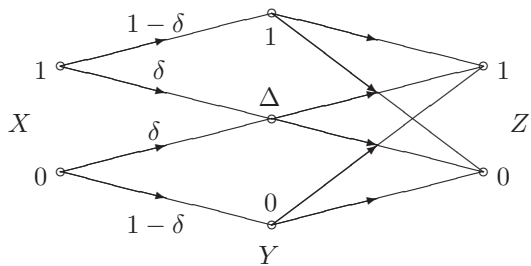

Fig. 2. Cascade of a binary erasure (BEC) and unknown channel

In most practical scenarios, only partial knowledge of the joint distribution is common. In a one-shot experiment as described, it is impossible to obtain a good estimate of $X$ based on $Z$ either with the direct Bayesian approach or with the role model approach. We will therefore assume that the experiment is repeated independently, so that the channel is driven with an independent and identically distributed (i.i.d.) random process $X_1, X_2, \ldots$. Since both channels are memoryless, their outputs are also i.i.d. random processes $Y_1, Y_2, \ldots$ and $Z_1, Z_2, \ldots$. For the direct Bayesian solution, we can adopt a frequentist approach to estimate the missing probabilities. Let us now concentrate on the role model approach. There are again two parameters that need to be optimized: $q_0 \stackrel{\text{def}}{=} Q_{X|Z}(0|0)$ and $q_1 \stackrel{\text{def}}{=} Q_{X|Z}(1|1)$. Since we do not know the conditional distribution $P_{Y|Z}$, we cannot, as in the previous example, simply determine $q_0$ by minimizing $\mathrm{E}D(P_{X|Y}||Q_{X|Z=0})$. Let us now consider the definition of the expected divergence (5), which we rewrite here for convenience

$$\mathrm{E}D(P_{X|Y}||Q_{X|Z}) = \sum_z \sum_y P(yz) D(P_{X|Y=y}||Q_{X|Z=z}).$$

Note that for any given $y$ and $z$, we can compute the divergence inside the expectation. We lack the knowledge about $P_{YZ}$ to be able to compute the expectation. Since the channels are memoryless, all the random processes involved are i.i.d. and therefore ergodic, and we can apply the law of large numbers, replacing the expectation by an average over time, i.e.,

$$\mathrm{E}D(P_{X|Y}||Q_{X|Z}) = \lim_{N \to \infty} \frac{1}{N} \sum_{i=1}^{N} D(P_{X_i|Y_i=y_i}||Q_{X_i|Z_i=z_i}), \tag{21}$$

with probability 1. Note that this is a mixed frequentist-probabilistic approach, since we are using the true *a posteriori* distribution $P_{X|Y}$ and optimizing a probabilistic model $Q_{X|Z}$, but applying the law of large numbers in order to compute the average. We never actually use measured frequencies as we would for the direct Bayesian approach.

Equation 21 gives us a method for approximating the expected divergence of interest for any given parameters $q_0$ and $q_1$ using a time average, but how do we proceed to optimize the choice of parameters in order to minimize the expected divergence? One solution is to use convex optimization, since the function and its average are convex in $q_0$ and $q_1$. We can apply any numerical convex optimization technique. Note that the partial derivatives of the convex function with respect to $q_0$ and $q_1$, which are required for some convex optimization techniques, can also be computed via time averaging, i.e.,

$$\begin{aligned}
\frac{\partial}{\partial q_0} \mathrm{E}D(P_{X|Y}||Q_{X|Z}) &= \frac{\partial}{\partial q_0} \sum_y P_{YZ}(y0) D(P_{X|Y=y}||Q_{X|Z=0}) \\
&= \frac{\partial}{\partial q_0} \sum_y P_{YZ}(y0) \left[ P_{X|Y}(0|y) \log_2 \frac{P_{X|Y}(0|y)}{q_0} \right. \\
&\qquad \left. + P_{X|Y}(1|y) \log_2 \frac{P_{X|Y}(1|y)}{1-q_0} \right] \\
&= \frac{\partial}{\partial q_0} \sum_y P_{YZ}(y0) \left[ -P_{X|Y}(0|y) \log_2 q_0 - P_{X|Y}(1|y) \log_2(1-q_0) \right] \\
&= \frac{1}{\log 2} \sum_y P_Z(0) P_{Y|Z}(y|0) \left[ \frac{P_{X|Y}(1|y)}{1-q_0} - \frac{P_{X|Y}(0|y)}{q_0} \right] \\
&= \lim_{N \to \infty} \frac{1}{N} \sum_{i=1}^{N} \frac{1(Z_i=0)}{\log 2} \left[ \frac{P_{X_i|Y_i}(1|y_i)}{1-q_0} - \frac{P_{X_i|Y_i}(0|y_i)}{q_0} \right], \quad (22)
\end{aligned}$$

with probability 1, where $1(.)$ denotes a function that is 1 when the equality holds and zero otherwise.

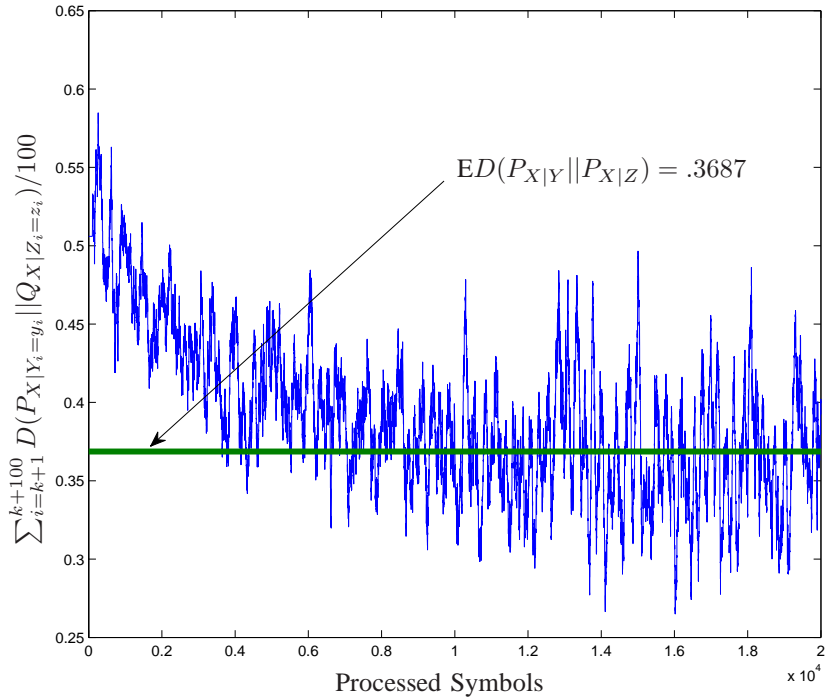

Fig. 3. Divergence measured through time averaging during the gradient descent optimization

We have tested this approach using a BEC with $\delta = 1/4$ and a channel from $Y$ to $Z$ with parameters $P_{Z|Y}(0|0) = 0.9$, $P_{Z|Y}(0|\Delta) = 0.7$, and $P_{Z|Y}(0|1) = 0.2$, all unknown to the receiver. All time averages are taken as moving averages over a window of 100 past observations, and we use a simple gradient search to optimize $q_0$ and $q_1$. The parameters are initialized as $q_0 = q_1 = 1/2$, and an optimization step is performed for every received symbol starting from the 101$^{\text{st}}$ symbol. Figure 3 shows the evolution of the average divergence throughout the experiment, and Figure 4 shows the evolution of the parameters $q_0$ and $q_1$. Note that once the divergence reaches its minimum, the variables have converged to $q_0 = .7234$ and $q_1 = .8182$, which corresponds to the correct *a posteriori* distribution $P_{X|Z}$ for the compound channel with the parameters we chose.

Note that this second example is a general test case for the use of the role model strategy, since we could use exactly the same method for any known discrete channel from $X$ to $Y$, including for example a finely quantized additive white Gaussian noise (AWGN) channel and any unknown discrete memoryless channel from $Y$ to $Z$. Since we have not yet extended the role model framework to continuous random variables, we cannot make any statement about its use for a continuous AWGN channel. The number of parameters to optimize will depend only on the alphabet size of $X$ and of $Z$ and not on the alphabet size of $Y$, e.g., when applied to a finely quantized binary-input AWGN channel, no matter how fine the quantization, we will still need to optimize only two parameters if the unknown channel has a binary output. For unknown channels with very large output alphabets, it may become impractical to optimize all the free parameters of the *a posteriori* distribution as we did in the example. In this case, a parametric model for $Q_{X|Z}$ can be adopted and the parameters of that model optimized. Note that, depending on the parametric model chosen, the problem may not remain convex.

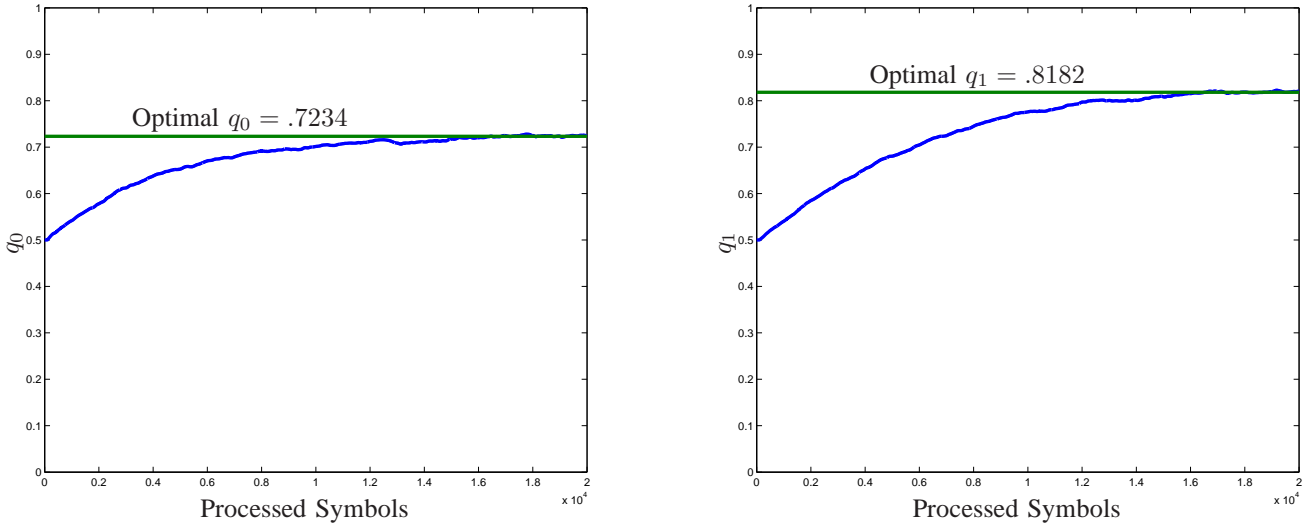

Fig. 4. Convergence of the parameters $q_0$ and $q_1$ in a gradient descent method

## V. APPLICATIONS

Markov chains are ubiquitous in all fields where probability theory is used, from physics to social sciences. However, not every scenario involving a Markov chain is a potential application for the role model strategy. A specific set of circumstances has to be fulfilled for the role model strategy to be useful.

In most practical applications, the joint distribution $P_{XYZ}$ is unknown, as was the case in our last example, and it is one of the aims of statistical modeling to "fill the gaps" and infer this distribution or one of its constituents $P_{X|Z}$ from the observed data.

When the following circumstances are fulfilled, the role model strategy becomes relevant:

- the direct solution $S_{\mathsf{d}} = P_{X|Z=z}$ is hard to obtain. This can have several reasons:
    1) the computation of $S_{\mathsf{d}}$ is too complex; or
    2) we have no access to $X$ when measuring the joint statistics of $X$ and $Z$. We need to construct an estimator for $X$ "blindly", e.g., without the ability to know or determine $X$ in a training phase of the estimator. In communications, this is often the case for models that must be adapted within the receiver without access to the transmitter; or
    3) the alphabet of $Z$ is too large to estimate a joint distribution of $X$ and $Z$ accurately using the available data;
- there exists a good statistical model for the joint distribution of $X$ and $Y$, allowing us to effectively compute $P_{X|Y=y}$ for any observation $y$. This may for example be a physical dependence model of $X$ on $Y$, or an existing model based on extensive statistical measurements that we are unable or unwilling to repeat for $X$ and $Z$;

- we are granted some access to $Y$ for training purposes. For example:
    1) $Y$ is available during an initial training phase; or
    2) $Y$ is available intermittently in a "now you see it, now you don't" fashion, and we can use the data points where both $Y$ and $Z$ are available in order to train an estimator that will work for data points where only $Z$ is available; or
    3) $Y$ is available for a sub-sample of a population of interest. This may be the case for example in applications where there is a cost associated with producing $Y$, so we choose to implement a few instances of $Y$ to train an estimator that will be applied to a much larger set of instances of $Z$.

There are numerous applications where this set of circumstances is fulfilled. We have successfully applied the role model strategy to the design of low-complexity decoders for low-density parity check (LDPC) codes, and the results are documented in [6]. Furthermore, a large number of so-called incomplete data problems correspond to the setup of Theorem 2 and the role model strategy can be applied to them. Let us consider a few hypothetical examples:

- zoologists wish to monitor the population of leopards in a national park by observing the leopard scats (a scientific word for faeces) found at various sites in the park. They received funding to conduct a DNA analysis of the scats over a period of 12 months. The DNA gives additional information that allows for a more precise estimation of the number of leopards present at each site. After that period, they must be able to continue monitoring the population of leopards without the extra information provided by the costly DNA analysis. This can be tackled by the role model approach: the estimator without DNA analysis can be trained to mimic the better estimator with DNA analysis over the year for which both are available;
- an internet search engine must constantly be refined so that its page ranking function remains in touch with user expectations and with the evolving internet. For some searches, contextual information about the user's past and related searches is available, making it easier to rank search results effectively. The role model approach can be applied to refine the ranking without contextual information to mimic the ranking with contextual information;
- a home insurance company computes the risk associated with each contract based on a number of input parameters, e.g., local crime rate, age of the property, price of the property, etc. When some of this information is missing, it is common for agents to provide ballpoint figures for the missing data. Instead, the role model approach can be used to refine a risk estimator based on partial data to mimic the estimates computed by the role model estimator based on the complete data.

Unfortunately, no internet search or insurance company has yet volunteered to let us perform measurements using their closely guarded data so we cannot yet report on the performance of the role model strategy in this context (anyone from these industries who happens to read this paper is cordially invited to contact the author...) The zoology example is based on a real world study [7] and, although the researchers would have been happy to volunteer their data for statistical testing, in reality they only found a single confirmed leopard, which is too sparse a sample to constitute a testing ground for the role model strategy. In the future, we will endeavour to find practical applications outside the field of receiver design where the role model strategy provides benefits.

## VI. Conclusion

We have introduced the "role model" strategy to build one statistical model by mimicking a better statistical model, and shown that this strategy is optimal under a Markov condition. We have shown examples of its use and discovered that, under certain circumstances, it yields a convex optimization problem that can be solved using well-known numerical techniques. In this paper, our aim was to isolate a theoretical result that we derived within our work and to show its applicability within a wider context. Its application to the design of low-complexity decoders for LDPC codes is described in detail in a separate paper [6].

A final remark concerns this paper's dedication and the "role model" relationship that underlies it: let $X$ be the problem of how to write a good paper, $Y$ be the skill and knowledge of the author's PhD advisor Jim Massey, and $Z$ be the skill and knowledge of this paper's author. The author did his best to apply the role model strategy while writing this paper. Whether the channel between $Y$ and $Z$ has non-zero capacity and whether $X - Y - Z$ form a Markov chain is for the reader to judge. On the other hand, further implications of this relationship could be investigated: Jim made it no secret during his lectures that his role model is Claude E. Shannon and that he does his best to emulate him. The problem of a multi-stage role model could be an interesting extension to the results presented here. Finally, should the associate editor handling this paper assign the review to Jim Massey himself, we would be in the realm of role models with feedback, another potentially interesting extension.

## Acknowledgment

The author benefited from discussions with David Declercq, Albert Guillén i Fàbregas, Gerhard Kramer, Alex Grant, and Maja Lončar during the preparation of this paper, and would like to express his warm gratitude to them.

## References

[1] G. Lechner and J. Sayir, "Improved sum-min decoding of LDPC codes," in *Proc. Int. Symp. Inf. Theory and Its App. (ISITA)*, Oct. 2004, pp. 997–1000.

[2] ——, "Improved sum-min decoding for irregular LDPC codes," in *Proc. Int. Symp. on Turbo Codes & Rel. Topics*, Apr. 2006.

[3] ——, "Improved sum-min decoding of irregular LDPC codes using nonlinear post-processing," in *NEWCOM-ACoRN Joint Workshop*, Sep. 2006.

[4] A. Ashikhmin, G. Kramer, and S. ten Brink, "Extrinsic information transfer functions: Model and erasure channel properties," *IEEE Trans. Inf. Theory*, vol. 50, no. 11, Nov. 2004.

[5] T. A. Cover and J. A. Thomas, *Elements of Information Theory*. John Wiley and Sons, 1991/2006.

[6] J. Sayir, "Information averaging for low complexity decoders," in preparation, to be submitted.

[7] I. Perez, E. Geffen, and O. Mokady, "Critically endangered Arabian leopards *Panthera pardus nimr* in Israel: estimating population parameters using molecular scatology," *Oryx, The International Journal of Conservation*, vol. 40, Jul. 2006.